\documentclass[aps,prb,twocolumn,noshowpacs,floatfix]{revtex4}
\usepackage{graphicx}
\usepackage{dcolumn}
\usepackage{bm}
\usepackage{amsmath, amsfonts}
\usepackage[breaklinks=true,colorlinks,citecolor=blue,linkcolor=blue,urlcolor=blue]{hyperref}
\def\be{\begin{equation}}
\def\ee{\end{equation}}
\def\nn{\nonumber}
\def\ber{\begin{eqnarray}}
\def\eer{\end{eqnarray}}
\begin{document}
%
\title{Time resolved transport properties of a $Y$-junction of 
Tomonaga-Luttinger liquid wires}
\author{Amit Agarwal}
\email{amitag@iitk.ac.in}
\affiliation{Department of Physics, Indian Institute of Technology, 
Kanpur 208 016, India}

\begin{abstract}
We study time resolved transport properties of a $Y$-junction composed 
of interacting one-dimensional quantum wires using a bosonization approach. 
In particular, we investigate the AC conductivity of the $Y$-junction 
formed from finite length Tomonaga-Luttinger liquid wires based on a 
plasmon scattering approach for injected charge pulses of arbitrary shapes.
In addition, we calculate the tunneling current and quantum noise of the 
$Y$-junction arising from point-like tunneling impurities at the junction, 
including finite temperature effects. Our results will be useful for 
designing nano-electronic quantum circuits, and for interpreting 
time-resolved experiments [H. Kamata {\it et. al.},  Nat. Nanotechnol.
{\bf 9}, 177 (2014)] in interacting wires and their junctions. 
\end{abstract}
\pacs{}
\maketitle
\section{Introduction}
One-dimensional (1D) quantum wires and the junctions of several 1D 
quantum wires are expected to be  important 
for potential applications as components in future nano-electronic devices.
Such 1D quantum wires with interacting electrons are described by 
the Tomonaga-Luttinger liquid (TLL) theory 
[\onlinecite{Tomonaga, Mattis, Haldane, Luttinger, delft, Giamarchi}], 
the low-energy excitations in 
which are the collective density oscillations. These density oscillations 
or plasmon modes, are markedly different from their counterparts - Landau's 
quasi-particle excitations, in higher dimensions described very successfully 
by the Fermi liquid (FL) theory [\onlinecite{Giovanni}]. This leads to 
unique physics in 1D, such as the spin-charge separation in which the 
spin and charge excitations propagate with different velocities 
[\onlinecite{Auslaender1, Jompol}], or the phenomena of charge 
fractionalization [\onlinecite{Safi, pham, Steinberg}]. Recently charge 
fractionalization has also been observed using time resolved measurements 
on coupled integer quantum Hall edge channels [\onlinecite{Kamata}]. 

In the present work, we investigate the time-dependent transport properties 
of multi-wire junctions, and a three-wire $Y$-junction in particular. These 
have already been realized 
experimentally in crossed single-walled carbon nanotubes 
[\onlinecite{fuhrer,terrones}]. Such $Y$-junctions with interacting 
quantum wires are also 
extremely `rich' from a basic physics viewpoint and continue to be explored 
very actively in the literature 
[\onlinecite{nayak, lal200, chen, chamon1, Meden_prb2005, das2, Giuliano, 
Bellazzini, Hou, agarwal_tdos, abhiram, dasrao, Rahmani, Feldman_PRB2011, 
wolfe1, Hou_prb2012}].
Earlier theoretical studies of $Y$-junctions have primarily focused on the 
fixed points of the junction, their stability analysis, and the associated DC 
conductivity. These studies either use the fermionic language and the weak 
interaction renormalization group (RG) approach [\onlinecite{lal200}], or 
the bosonic and conformal field theory language [\onlinecite{chamon1, das2}], 
or other numerical 
methods such as the functional RG [\onlinecite{Meden_prb2005}].
A comprehensive study of the fixed points of the $Y$-junction formed from 
spin-less interacting electrons, and the DC conductance, was carried out in 
Ref.~[\onlinecite{chamon1}]. This was later extended to include spin-ful 
electrons giving a much richer phase diagram in the parameter space of charge 
and spin interactions [\onlinecite{Hou}], and to account for different 
interaction strengths in different wires [\onlinecite{Hou_prb2012}]. 

\begin{figure}[t]
\begin{center} 
\includegraphics[width=.98 \linewidth]{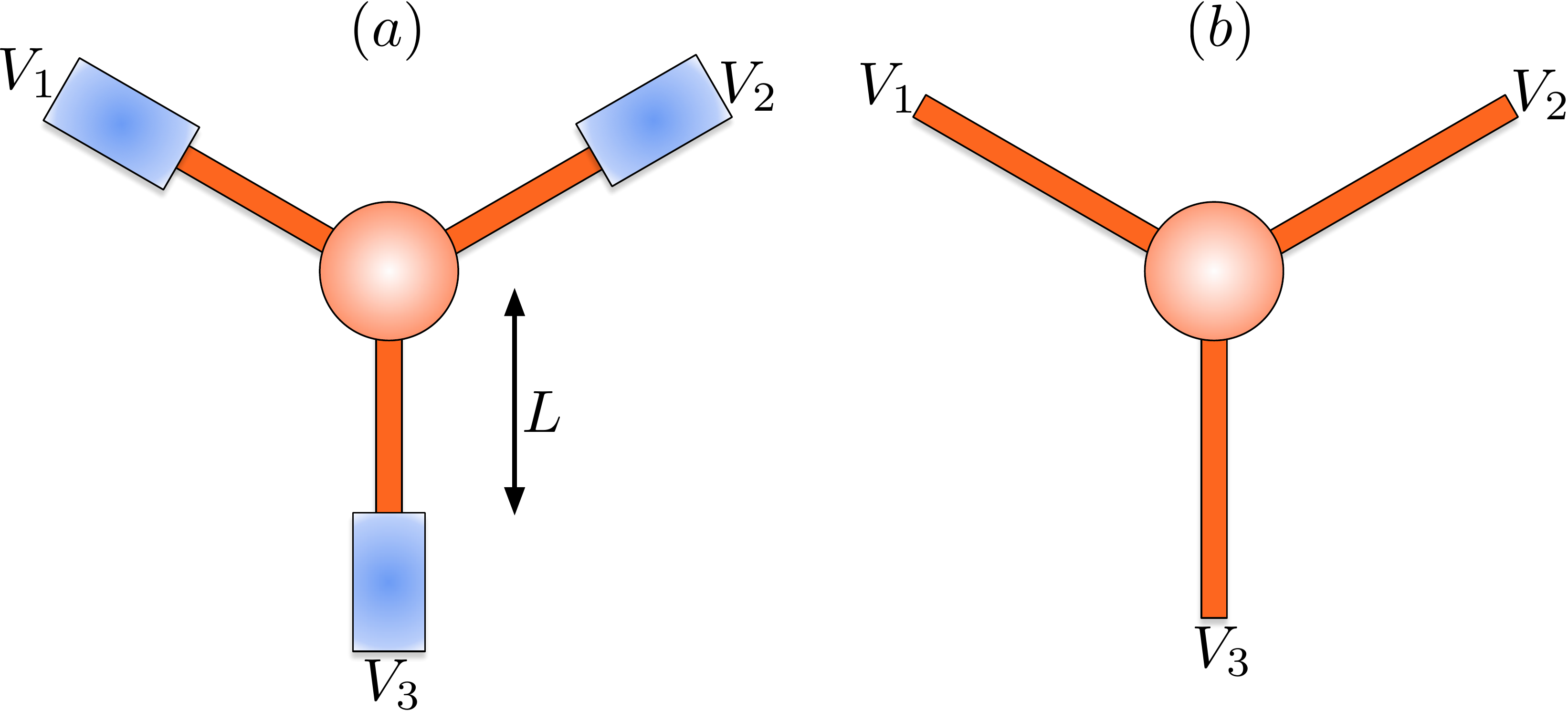}
\end{center}
\caption{Schematic of a $Y$-junction composed of TLL wires. Panel (a) shows 
a $Y$-junction of finite length TLL wires (red) connected to Fermi liquid 
leads at the ends (blue) with different applied voltages.  Panel (b) displays 
a $Y$-junction of infinite TLL wires.}
\label{fig0} 
\end{figure}

Time-dependent transport properties of 1D TLL wires have also been studied 
earlier. Quantum noise for an infinite TLL wire with point-like tunneling 
impurity, around the `connected' fixed point of a two wire junction was studied in 
Ref.~[\onlinecite{chamon_prb1995}].
The AC conductivity of a clean finite length TLL wire was calculated in 
Refs.~[\onlinecite{Safi},\onlinecite{Berg}]. This has been recently 
generalized to include arbitrary wave packet shapes of the incident 
current in Ref.~[\onlinecite{Perfetto2014}]. Comparatively, the 
time-dependent transport properties of $Y$-junctions, have drawn much less 
attention in the literature and it is the aim of this article to rectify this. 

In this article we study the AC conductivity, the tunneling current and 
quantum noise (including shot noise and Josephson noise) of a $Y$-junction 
tuned to a dissipation-less fixed point with spin-less TLL wires. We consider 
both time-reversal symmetry (TRS) preserving and TRS violating junctions and 
use the single-parameter description of the dissipation-less fixed points of 
the junction given in Refs.~[\onlinecite{das2},\onlinecite{agarwal_tdos}]. 
Our analysis may be useful for interpreting time-resolved experiments 
[\onlinecite{Kamata}] in multi-wire junctions and for designing 
nano-electronic quantum circuits [\onlinecite{Jezouin}].

This article is organized as follows. In Sec.~\ref{secII} we discuss the 
details of the three-wire $Y$-junction and show that both the Coulomb 
interactions in the wire and the `scattering' boundary conditions at the 
junction, can be treated using bosonization with delayed evaluation of the 
boundary conditions [\onlinecite{chamon1}].
In Sec.~\ref{secIII}, we calculate the AC 
conductivity of the $Y$-junction formed from finite 
length TLL wires which are connected to FL leads --- see Fig.~\ref{fig0} (a).
We also reproduce the known results for a two-wire junction, and the 
DC conductivity as a limiting case of our calculations.
In Sec.~\ref{secIV}, we calculate the tunneling current and quantum noise 
at the junction with infinite TLL wires [see Fig.~\ref{fig0} (b)], 
in the presence of point-like tunneling impurities at the $Y$-junction 
tuned to a dissipation less fixed point. Finally
we summarize our findings in Sec.~\ref{summary}.

\section{Bosonization of the junction -- delayed evaluation of the boundary 
condition} 
In this section we review the technique of bosonization for the wire, 
and subsequently the parametrization of the dissipation-less fixed 
points at the junction.
\label{secII}
 \subsection{Bosonization of the wires} 

To  model a junction of multiple  wires,  let us assume that N semi-infinite wires meet at a junction. The wires are modeled as spin-less
TLL on a half-line and are parametrized by coordinates $x_i, i =1,2, . . . ,N$   such that ($x_i > 0$).
We use a folded basis to describe the junction, {\it i.e.~}we choose a convention that for all the wires, $x_i = 0$ at the junction and $x_i$ increases from $0$ as one goes outwards from the wires.
We denote the incoming and outgoing single electron wave functions on wire $i$ by $\phi_{iI}$ and $\phi_{iO}$ respectively, which in turn are proportional to plane waves $\exp{[-i k( x_i + v t)]}$ and  $\exp{[i k( x_i - v t)]}$ respectively, for a given wavenumber $k > 0$ and velocity $v$.
For simplicity of analysis, 
we consider all the semi-infinite spin-less TLL wires to have the same short ranged 
electron-electron ({\it e-e}) interaction strength and Fermi velocity. 
%
%
%

The spin-less electron field on each wire can be expressed as $\psi (x) = 
\psi_{\rm I} (x) + \psi_{\rm O}(x) $ where the incoming/outgoing fermionic 
fields $\psi_{\rm I /\rm O} $ can be bosonized [\onlinecite{delft}] as 
\ber \label{BI}
 \psi_{\rm I} (x) &=&\frac{1}{\sqrt{2 \pi \alpha}} F_{\rm I}~
e^{2i\pi N_{\rm I} (x +  vt)/L}~e^{ - i k_{\rm F} x + i 
\phi_{\rm I}(x)} ~, \nn \\ 
 \psi_{\rm O} (x) &=& \frac{1}{\sqrt{2 \pi \alpha}} F_{\rm O}~
e^{2i\pi N_{\rm O} (x -  vt)/L}~e^{  i k_{\rm F} x + i 
\phi_{\rm O}(x)} ~.
\eer
Here $F_{\rm I}$ and $F_{\rm O}$ are Klein factors 
for the incoming and outgoing electrons respectively, $k_{\rm F}$ is the 
Fermi momentum, and $\alpha$ is the inverse ultraviolet (short distance) cut-off. 
$N_{\rm I}$ and $N_{\rm O}$ count the number of incoming and outgoing chiral 
particles with respect to the filled Fermi sea. The fields $\phi_{\rm I}(x)$ 
and $\phi_{\rm O}(x)$ are the incoming (left moving) and the outgoing 
(right moving) 
chiral bosonic fields in each wire and can be expressed in terms of the 
bosonic creation and destruction operators as, 
\be \phi_{\rm O/I} \equiv \sum_{q>0}\frac{1}{\sqrt{n_q}} (b_{q{\rm O/I}}
e^{\pm iqx} + b^\dagger_{q{\rm O/I}}e^{\mp iqx}) e^{-\alpha|q|/2}~. \ee

The Lagrangian of the system is given by ${\cal L} = {\cal L}_0+ 
{\cal L}_{\rm int}$ where ${\cal L}_0$ describes free electrons in the wire, 
and is given by 
\ber {\cal L}_0 &=& \frac{1}{4\pi} \sum_{i=1}^N \int_0^L dx ~ \Big[ \partial_x 
\phi_{i{\rm O}}~ (- \partial_t - v \partial_x) ~\phi_{i{\rm O}} \nonumber \\ 
& +& \partial_x \phi_{iI} ~(\partial_t - v \partial_x) ~\phi_{iI}) \Big]~, 
\label{lag11} \eer
where $v$ denotes the Fermi velocity which we take to be same in all the 
wires and $i$ is the wire index. The corresponding incoming and outgoing 
density and current fields in each wire are given by 
\ber \label{eq:density}
\rho_{iO} = \frac{\partial_x \phi_{iO}}{2\pi}+ \frac{N_{iO}}{L}~, & & 
J_{iO} = -\frac{\partial_t \phi_{iO}}{2\pi} - \frac{N_{iO}}{L} ~,\nn \\
\rho_{iI} = - \frac{\partial_x \phi_{iI}}{2\pi}+ \frac{N_{iI}}{L}~, & & J_{iI} 
= ~\frac{\partial_t \phi_{iI}}{2\pi} - \frac{N_{iI}}{L}~. 
\eer
We emphasize here that the second term in the expressions for density and current 
arise from excessive number of incoming and outgoing fermions with respect to the ground state (filled Fermi sea), and can be controlled by 
applying an external DC voltage in each TLL wire. These terms will be very useful in Sec.~\ref{secIV}, where we apply different DC bias voltage on 
each of the three wires. However for calculating the AC conductivity in Sec. \ref{secIII}, only the first term of the 
current expression (temporal derivative of the fluctuating fields) is needed since the average DC voltage is zero in all the wires, and we will use the notation,  
\be j_{iO} =  -\frac{\partial_t \phi_{iO}}{2\pi}~,~~~ {\rm and }~~~ j_{iI}  = ~\frac{\partial_t \phi_{iI}}{2\pi}~,
\label{eq:j} 
\ee
in Sec.~\ref{secIII}

For a short range {\it e-e} interaction between the two chiral modes in the wire,
the term in the Lagrangian for each wire $i$ is of the form 
\be {\cal L}_{int}^i = \frac{\lambda}{4\pi} ~ \int_0^L dx ~ \partial_x 
\phi_{iI} ~ \partial_x \phi_{iO}~, \label{lag20} \ee
where $\lambda$ is the {\it e-e} interaction strength (positive for repulsive 
interactions) with the dimensions of velocity. Note that for each of the wires 
described by Eqs.~\eqref{lag11} and \eqref{lag20}, the effective TLL velocity
and the effective TLL interaction strength are given by 
\be \label{eq:v}
\tilde{v} = \sqrt{v^2 -\lambda^2/4}~, ~~~~{\rm and}~~~~g = \sqrt{ 
\frac{v - \lambda /2}{v + \lambda/2}}~. \ee 

\subsection{Bosonization of the junction}

To describe the junction uniquely, we need to impose an appropriate boundary 
condition on the fields at the junction, {\it i.e.~}at $x=0$. Following 
standard procedure [\onlinecite{chamon1, das2, agarwal_tdos}], the incoming and 
outgoing currents, and consequently the bosonic fields, are related at the 
junction by a current splitting matrix ${\mathbb{M}}$, {\it i.e.~} $j_{Oi} = 
\sum_j {\mathbb{M}}_{ij} ~ j_{Ij}$, which leads to $\phi_{Oi} = \sum_j
{\mathbb{M}}_{ij} ~\phi_{Ij}$. Here we have ignored an integration constant 
which plays no (physical) role in the computation of the Green's functions of 
the fields and consequently 
on the scaling dimensions of various operators. In order to
ensure that the matrix ${\mathbb{M}}$ represents a fixed point of the
theory, the incoming and outgoing fields must satisfy appropriate bosonic
commutation relations; this restricts the matrix ${\mathbb{M}}$ to be
orthogonal. Scale invariance or conformal invariance imposes the same 
constraints of orthogonality [\onlinecite{dasrao}] on ${\mathbb{M}}$. 
The constraint 
of orthogonality also implies that there is no dissipation in the system 
[\onlinecite{agarwal_prb2009}]. In addition, to ensure current conservation 
at the junction, its rows (or columns) have to add up to unity. 

Since $\phi_{O}$ and $\phi_{I}$ are interacting fields, we need to perform 
a Bogoliubov transformation on them, 
\be \phi_{O/I}=\frac{1}{2{\sqrt g}}\left[(1 +g)\tilde{\phi}_{O/I} + 
(1 - g){\tilde {\phi}_{I/O}} \right], \ee 
to obtain the corresponding `free' outgoing and incoming ($\tilde\phi_{O/I}$)
chiral fields, which satisfy the `free' field commutation relations: 
$[\tilde{\phi}_{O/I}(x,t),\tilde{\phi}_{O/I}(x',t)]=\pm i\pi
{\rm sign} (x-x')$, where the sign function is defined as $\text{sign}(x) = 
1,0,-1$ for $x>0, ~x=0$ and $x<0$ respectively. 
However, unlike the usual Bogoliubov transformation in
the bulk, here we also need to consider the effect of the junction
matrix ${\mathbb{M}}$ relating the interacting incoming and outgoing fields 
[\onlinecite{das2}], which leads to a `Bogoliubov transformation' of the 
matrix: 
${\mathbb{M}} \to \widetilde{\mathbb{M}}$. Qualitatively, ${\mathbb{M}}$ is 
related to tunnelings between the different wires and tunneling in each wire, 
at a dissipation-less junction. The Bogoliubov transformed matrix 
$\widetilde{\mathbb{M}}$ which relates the `free' incoming and outgoing fields,
$\tilde{\phi}_{Oi} (x) = \sum_j ~ \widetilde{{\mathbb{M}}}_{ij}
~\tilde{\phi}_{Ij} (-x)$, is given by 
\be \widetilde{{\mathbb{M}}} = \left[(1+g){\mathbb{I}}-(1-g){\mathbb{M}}
\right]^{-1} \left[(1+g){\mathbb{M}}-(1-g)\mathbb{I}\right]~. \ee 
We emphasize that this description is valid for a dissipation-less junction of 
any number of interacting one-dimensional wires. 

For the case of a two-wire junction, there are only two classes of 
orthogonal matrices: a rotation matrix whose determinant is $1$ and a 
reflection matrix whose determinant is $-1$. The constraint that the columns 
(or rows) add up to one, imply that there is only one matrix in each class. 
These are given by
\be \left(\begin{array}{cc}
1 & 0 \\
0 & 1 \\ \end{array}\right)~, \quad{\rm and} \quad
\left(\begin{array}{cc}
0 & 1 \\
1 & 0 \\ \end{array}\right)~, 
\ee 
which corresponds to the cases of the `dis-connected' and the `connected' 
fixed points of a two-wire junction, respectively.
In what follows, we focus on a three-wire $Y$-junction.

A detailed study of the three-wire spin-less TLL junction using bosonization 
and boundary conformal field theory can be found
in Refs.~[\onlinecite{chamon1,dasrao}]. 
In particular, for a three-wire charge-conserving junction all current splitting orthogonal matrices ${\mathbb M}$ whose rows add up to one can be parametrized by a single continuous parameter $\theta$, and are divided into 
two classes on the basis of TRS: 
$\det {\mathbb{M}}_1 = 1$, and $\det {\mathbb{M}}_2=-1$. These two classes 
of matrices are explicitly given by
\be \label{m1} {\mathbb{M}}_1 = \left(\begin{array}{ccc}
a & b & c \\
c & a & b \\
b & c & a \end{array}\right), \quad {\mathbb{M}}_2 =
\left(\begin{array}{ccc}
b & a & c \\
a & c & b \\
c & b & a \end{array}\right)~, 
\ee 
where, $a=(1+2\cos\theta)/3$, $b=(1-\cos \theta+ \sqrt{3} \sin \theta)/3$, and
$c=(1-\cos\theta -\sqrt{3} \sin\theta)/ 3$.
This gives us an explicit single parameter characterization of the two
families of fixed points; any fixed point in the
theory can now be identified in terms of $\theta$, with the fixed
points at $\theta=0$ and $\theta=2 \pi$ being identical. 

Note that the current splitting matrix ${\mathbb{M}}$ preserves TRS, 
only if it is symmetric. Thus the junction current splitting matrices 
belonging to the ${\mathbb{M}}_2 $ class, represents an asymmetric class
(in wire indices) of fixed points for systems with TRS.
The ${\mathbb{M}}_1$ class represents a $Z_3$ symmetric (in the wire 
index) class of fixed points and generally denotes systems with broken 
TRS, which can arise, for instance, due to a magnetic field at the 
junction (assuming a finite cross-sectional area). In 
the ${\mathbb{M}}_1$ class of fixed points, only two points given by 
$\theta = 0, \pi$, at which the asymmetry producing $\sin \theta$ term 
vanishes, are TRS invariant.
For the ${\mathbb{M}}_1$ class, $\theta = \pi$, or $[a,b,c] = [-1/3, 2/3, 
2/3]$, corresponds to the so called Dirichlet fixed point ($D_P$). The 
disconnected fixed point ($N$), where there is no tunneling between any 
pair of wires, is given by $\theta=0$ ({\it i.e.~}$[a,b,c] = [1,0,0]$). 
The case of $\theta = 2\pi/3$ ({\it i.e.~} $[a,b,c] = [0,1,0]$) and $4\pi/3$ 
correspond to the chiral $\chi_{-}$ and $\chi_{+}$ fixed points respectively, following 
the notation of Ref.~[\onlinecite{chamon1}]. 

Note that the ${\mathbb{M}}_2$ class of fixed point matrices has the 
interesting property that $({\mathbb{M}}_2)^2=\mathbb{I}$. As a consequence
$\widetilde{{\mathbb{M}}}_2={\mathbb{M}}_2$, which implies that both the
interacting and the free fields satisfy identical boundary
conditions at the junction. 
This is not true for the ${\mathbb{M}}_1$ class of fixed point matrices, but 
the matrix $\widetilde{{\mathbb{M}}}_1$ still has the same form as the matrix
${{\mathbb{M}}}_1$ with the corresponding parameters given by $\tilde a =
{(3g^2-1 + (3g^2+1)\cos{\theta})}/{\delta}$ and $\tilde b / \tilde c
= {2(1 - \cos{\theta} \pm \sqrt{3} g \sin{\theta} )}/\delta$, where
$\delta={3(1+g^2+ (g^2-1)\cos{\theta})}$. Note that the matrices
$\widetilde{{\mathbb{M}}}_1$ are non-linear functions of the TLL
parameter $g$, while the matrices $\widetilde{{\mathbb{M}}}_2$ are
independent of $g$. This will have non-trivial manifestations for physical 
observables ({\it e.g.}~quantum noise, tunneling current etc. --- see 
Sec.~\ref{secIV}), when we consider a junction slightly 
away form the fixed points, as scaling dimensions of operators switched on 
perturbatively around the ${\mathbb{M}}_1$ class will generally be non-linear 
functions of $g$. 
On the other hand, for the ${\mathbb{M}}_2$ class of fixed points, the
scaling dimensions of operators, will always be linear functions of $g$. 

Having characterized the junction, we now proceed to study the AC 
conductivity of a $Y$-junction formed from finite length TLL wires, 
connected to FL leads --- see Fig.~\ref{fig0} (a).

\section{AC conductivity} 
\label{secIII}

In this section, we consider an incident charge wave packet originating in 
the FL lead connected to one of the TLL wires, say $i$, and its 
consequent motion after undergoing charge fractionalization at the FL-TLL 
boundaries and at the junction. This will also allow us to calculate the 
low frequency AC current splitting matrix $\mathbb{S}$, which relates the 
complex amplitudes of the incoming AC currents, to the complex amplitudes of 
the outgoing current, in the linear response regime. Such a time resolved 
measurement of an incident wave packet in a TLL wire of integer quantum Hall 
edge channels, was recently used to identify a single charge fractionalization 
event [\onlinecite{Kamata}]. In our language this correspond to a 
two-wire junction tuned to be at the `connected' fixed point 
(effectively a single finite length TLL wire connected to FL leads).

In the DC limit, {\it i.e.}~$\omega \to 0$, all signatures of charge 
fractionalization are lost and $\mathbb{S} \to \mathbb{M}$, which is the 
non-interacting current splitting matrix for a junction with finite TLL wires 
connected to the FL leads [Fig.~\ref{fig0} (a)].
For a junction with TLL wires extending to infinity, it simply reduces to 
the interacting current splitting matrix at the junction for all 
frequencies: $\mathbb{S} \to\widetilde{\mathbb{M}}$, since there is no 
FL-TLL interface. However, for finite length wires at finite frequencies, 
$\mathbb{S}$ depends on the fixed point, 
the strength of the {\it e-e} interaction, and the length of the TLL wires $L$, 
and it carries the signature of charge fractionalization events at the 
FL-TLL boundary.

In our model of the junction, there is no mechanism of power 
dissipation. Thus the average over one oscillation cycle of 
the incoming energy must be equal to the average outgoing energy per cycle. 
This imposes the constraint of unitarity on the 
${\mathbb S}$ matrix, which also serves as a useful check for our 
calculations. Also note that we are considering all  three wires to 
have the same Fermi velocity and {\it e-e} interaction strengths, and these are 
connected at the junction described by 
boundary conditions which are cyclic in nature. Thus we expect to have only 
a few independent coefficients in 
${\mathbb S}$ which should also appear in a cyclic manner. 

Before discussing the solution of the generalized plasmon scattering problem 
[\onlinecite{Perfetto2014}] at the junction, we emphasize that this 
calculation is valid only in the linear response regime and only for AC 
frequencies which do not breach the linearization regime for each TLL wire, 
{\it i.e.~}$ \omega < v/\alpha$. 
Also note again that we use a folded basis for describing the junction 
such that all the wires go from $x = 0$ to $x = \infty$ and the 
junction lies at $x = 0$.

The time evolution of the `injected' wave packet is given by the coupled 
equation of motion (EOM) for the expectation value of the incoming 
($\phi_{iI}$) and outgoing ($\phi_{iO}$) fields in wire $i$, which are 
governed by the Lagrangian given in Eqs.~(\ref{lag11}) and~(\ref{lag20}). 
The EOM are 
\ber \partial_x\left[\partial_t \phi_{iI} - v \partial_x \phi_{iI} + 
\frac{\lambda}{2} \partial_x \phi_{iO}\right] &=& 0~, \label{EOM1}
\\
\partial_x\left[\partial_t \phi_{iO} + v \partial_x \phi_{iO} - 
\frac{\lambda}{2} \partial_x \phi_{iI}\right]&=&0~. \label{EOM2} \eer
Let us now consider an electronic wave packet incident on TLL wire $i$ from 
the FL lead. The incoming bosonic field $\phi_I (x,t)$ in the FL lead ($x>L$), can 
be expressed, in terms of scattering states of energy $\omega = v q $, by the 
following relation 
\be \phi_{iI} (x,t) = \int_{-\infty}^{\infty} \frac{dq}{2\pi} \phi_{i I} (q)  
e^{- i( q (x-L) + \omega t)} ~. \label{eq:phiiI}  \ee 
Here $\phi_{i I } (q)$ is specified by the Fourier transform $\rho_{i I} (q)$ of the 
incident charge density in wire $i$, $\rho_{iI}(x,t=0) $, by the relation 
$\phi_{iI}(q) = \frac{2 \pi}{i q}\rho_{i I } (q)$ --- see Eq.~\eqref{eq:density}. The 
extra factor of $e^{i q L}$ in the above equation just shifts the position of 
the origin of the axis in the FL leads, and it simplifies the calculations 
below. The outgoing bosonic scattering state in the FL lead of wire $j$ due to in injected 
state in wire $i$ only, 
is given by 
\be \phi_{jO}^{(i)} (x,t) = \int_{-\infty}^{\infty} \frac{dq}{2\pi} \phi_{j O }^{(i)} (q)
e^{ i( q(x-L) - \omega t)} ~,  \label{eq:phijO}
 \ee
where the outgoing amplitude in the momentum space is related to the 
incoming amplitude via the elements of the AC current splitting matrix: 
\be
\phi_{jO}^{(i)}(q)  = s_{ji}(q) \phi_{iI} (q)~, 
\ee
with $s_{ji}$ denoting the matrix elements of ${\mathbb S}$ and $q = \omega/v$ .
We emphasise here that we are considering all the wires to have the same Fermi velocity. 
In the case of bosonic states being incident in all the wires, the total outgoing bosonic field 
gets contribution from all the incoming fields and it is explicitly given by $\phi_{jO} (x,t) = \sum \phi_{jO}^{(i)} (x,t) $, 
or equivalently, 
\be 
\phi_{jO} (x,t) = \sum_{i} \int_{-\infty}^{\infty} \frac{dq}{2\pi} s_{ji}(q) \phi_{i I } (q)
e^{ i( q(x-L) - \omega t)} ~.\ee

 If the elements $s_{ij}$ of $\mathbb{S}$ are known, then the 
total time-dependent density ($\rho = \rho_I + \rho_O$) and the total 
outgoing current ($j = j_I + j_O$) in the FL of wire $j$, due to an 
incoming wave packet in wire $i$, is given by 
\be \rho_j^{(i)} (x,t) = \int_{-\infty}^{\infty} \frac{dq}{2\pi} \rho_{iI} (q) e^{-i 
\omega t} \left[e^{-i q (x-L)} \delta_{ij}+ s_{ji} e^{i q(x-L)}\right], \ee
and 
\be j_j^{(i)} (x,t) = v \int_{-\infty}^{\infty} \frac{dq}{2\pi} \rho_{i I}(q) e^{-i 
\omega t} \left[ -e^{-i q (x-L)} \delta_{ij}+ s_{ji} e^{i q(x-L)}\right]. \ee 

In the TLL wire region ($x<L$), the incoming and outgoing fields, corresponding to a situation when there is only an incoming filed in wire $i$, are given by 
\be \label{eq:ansatz}
\phi_{j\frac{I}{O}}^{(i)} (x,t) = \int_{-\infty}^{\infty} \frac{dq}{2\pi} \phi_{i I} (q) 
e^{ -i \omega t} \left(a_{j\frac{I}{O}}^{(i)} e^{ -ikx} + b_{j\frac{I}{O}}^{(i)} e^{ i kx}
\right). \ee 
Here the AC frequency $\omega = \tilde{v} k$, where $\tilde{v}$ is the 
renormalized Fermi velocity in the TLL region and is given by Eq.~\eqref{eq:v}. 
Note that in Eq.~\eqref{eq:ansatz} above, $q$  is the wave-vector in the noninteracting FL leads, and 
$k$ denotes the wave vector in the interacting TLL region  for the fixed incoming energy $\omega$ and they are related to each other via the equation $k = v q /\tilde{v}$. 

We now proceed to solve the `plasmon scattering' problem and obtain the 
elements of $\mathbb{S}$. 
Let us consider an incoming current (from FL lead) only in wire $1$.
The continuity of the incoming and the outgoing currents at $x=L$, [using Eqs.~\eqref{eq:phiiI}-\eqref{eq:phijO}, and Eq.~\eqref{eq:ansatz},  in Eq.~\eqref{eq:j}]
gives the following equations in each wire (six in all) 
\ber & & a_{iI}^{(1)}~ e^{-i k L} ~+~ b_{iI}^{(1)}~ e^{i k L} ~=~ \delta_{i1}~, \quad 
{\rm and} \\
& & a_{iO}^{(1)}~e^{-i k L} ~+~ b_{iO}^{(1)}~ e^{i k L} ~=~ s_{i1} ~, \eer
where $s_{i1}$ are the elements of the
first column of ${\mathbb S}$, and the superscript is used to indicate that the incoming current is in wire $1$. Within the TLL region ($x<L$), 
substituting Eq.~\eqref{eq:ansatz} in Eqs. \eqref{EOM1}-\eqref{EOM2}, 
gives the following set of equations for each wire (six in all):
\ber & & 2 (\omega - v k )~a_{iI}^{(1)}~+~ k \lambda ~a_{iO}^{(1)}~=~0 ~,~~~{\rm and} \\
& & 2 (\omega - v k )~b_{iI}^{(1)}~-~k \lambda ~b_{iO}^{(1)}~=~0 ~, \label{eq:24}\eer
in addition to the consistency condition, $\omega=\tilde v k$, with 
$\tilde{v}=\sqrt{v^2 - \lambda^2/4}$. Besides 
these the boundary condition at the junction ($x=0$) is given by the field 
(current) splitting matrix ${\mathbb M}$ as
\be a_{iO}^{(1)} ~+~ b_{iO}^{(1)} ~=~ \sum_j~{\mathbb{M}}_{ij}~ ( a_{jI}^{(1)} ~+~ b_{jI}^{(1)}) ~.
\ee
Solving these 15 equations simultaneously gives us the three elements of 
the first column of the AC current splitting matrix. Repeating this 
calculation for the case with an incoming unit current in the other wires 
will give us the elements in the other two columns.

%
\begin{figure}[t]
\begin{center} 
\includegraphics[width=1.0 \linewidth]{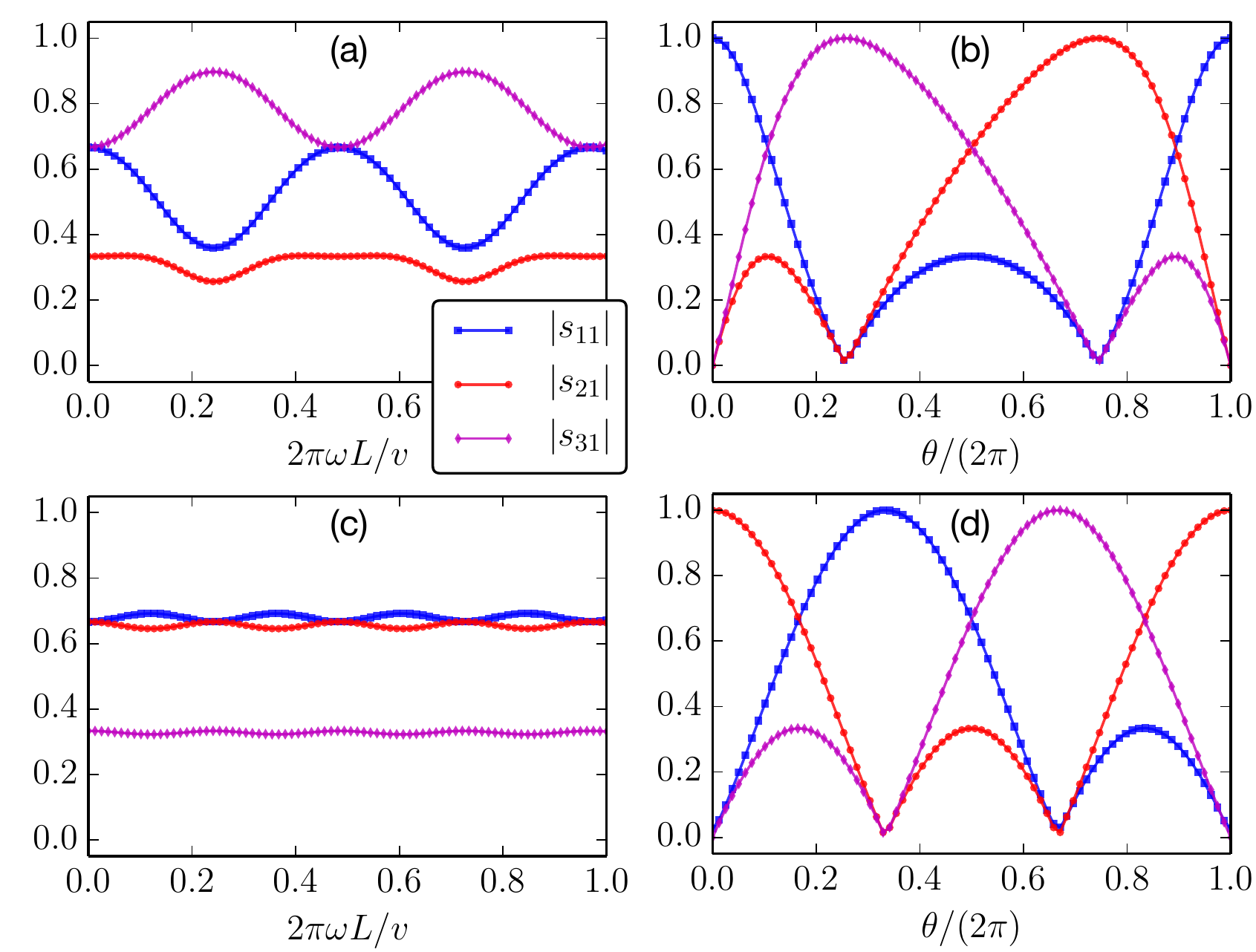}
\end{center}
\caption{The amplitudes of the AC current splitting matrix for (a) the 
${\mathbb{M}}_1$ class as a function of the frequency $\omega$ at the fixed 
point parameterized by $~\theta = \pi/3$, (b) the ${\mathbb{M}}_1$ class as 
a function of $\theta$ for $\omega L/v=\pi/2$, (c) the 
${\mathbb{M}}_2$ class as a function of $\omega$ with $~\theta = \pi/3$, and
(d) the ${\mathbb{M}}_2$ class as a function of $\theta$ for $\omega 
L/v=\pi/2$. 
The curves marked by blue squares, red circles, and magenta diamonds, represent 
$|s_{11}|,~|s_{21}|,~|s_{31}|$ respectively in all the panels. 
The {\it e-e} interaction strength is chosen to be $\lambda = 0.5 v$~ (which gives 
$\tilde{v}=0.97 v$ and $g = 0.88$). 
\label{fig1} }
\end{figure} 

\subsection{TRS preserving (${\mathbb M}_2$) fixed points}

Let us first consider the TRS preserving systems, {\it i.e.~}
$Y$-junctions with the ${\mathbb{M}}_2$ class of fixed points. Following the 
procedure described above, we calculate the AC current splitting matrix 
${\mathbb S}$, which has only six independent components. These are are 
given by 
\ber 
s_{11} &= & \frac{1}{\xi} \left[ 2~ \tilde v ~b - i ~\lambda ~\sin (2k L) \right]~, \label{s11_a}\\
s_{22}&=& \frac{1}{\xi} \left[ 2 ~\tilde v ~c - i ~\lambda ~\sin (2k L) \right]~ , \\
s_{33}&=& \frac{1}{\xi} \left[ 2~ \tilde v ~a - i~ \lambda~ \sin (2k L) \right] ~,\\ 
s_{21}&=& \frac{2}{\xi} ~\tilde{v}~a ~, \label{s21_a} \\ 
s_{31}&=& \frac{2}{\xi} ~\tilde{v} ~c ~, \\
s_{32}&=& \frac{2}{\xi} ~ \tilde{v}~b ~,  \label{s32_a}
\eer
where $a$, $b$ and $c$ are defined below Eq.~\eqref{m1},  and finally 
\be \xi = 2 \left[\tilde{v}~\cos (2k L) - i v~\sin (2 k L) \right] ~. \label{xi}\ee
We note again that the omega dependence of $\mathbb{S}$, 
appears in Eqs.~\eqref{s11_a}-\eqref{xi}, via $k$ which is defined after Eq.~\eqref{eq:24}.

The other elements of $\mathbb{S}$ are given by $s_{12}=s_{21},~
s_{13}=s_{31}$ and finally$~s_{23}=
s_{32}$. Note that the three off-diagonal 
elements and the three diagonal elements have a very similar structure and 
differ only due to the different corresponding element in the current 
splitting matrix ${\mathbb{M}}_2$ at the junction.

To get some physical insight for the form of ${\mathbb S}$, let us consider 
the specific case of a junction with $\theta = 0$ in the ${\mathbb{M}}_2$ 
class, {\it i.e.}
$[a, b, c] = [1,0,0]$, which parameterizes the case of wires $1$ and $2$ being 
directly connected effectively becoming one wire of length $2L$, and the wire 
$3$ being completely disconnected from the other two. 
For this case, Eq.~\eqref{s11_a} and Eq.~\eqref{s21_a} reduce to, 
\ber \label{chk1} s_{11} &=& \frac{ -i (\lambda/2) 
\sin{2 k L}}{\tilde{v}~\cos (2k L) - i v~\sin (2 k L) }~, \\
\label{chk2} s_{21} &=& \frac{ \tilde v}{\tilde{v}~\cos (2k L) - 
i v~\sin (2 k L)}~. \eer
As a check of our calculations we note that Eqs.~\eqref{chk1} and \eqref{chk2} 
are identical to the set of equations given in Eq.~(17) of 
Ref.~[\onlinecite{agarwal_ac}], which were derived 
for counter-propagating quantum Hall edge states which interact with each 
other. Furthermore this simpler case can also be 
derived by considering a step-like variation of the interaction strength, 
{\it i.e.}~$g(x) = g$ for $-L<x<L$, and $g(x) =1$ otherwise, in an interacting
1D wire [\onlinecite{Berg, Safi}]. Consider an electronic wave packet incident 
on the interacting region from the non-interacting region. Fractionalization 
of charge [\onlinecite{pham}]
in the interacting region, implies the reflection of fractional 
charge $q^* = r_0 e$, where $r_0 = (1-g)/(1+g) $ and transmission of a 
fractional charge $q^* = t_0 e$ into the interacting region, where $t_0 = 
2 g/(1+g)$. Other reflection and transmission coefficients for a single 
impact are given by $r_0' = -r_0$, and $t_0' = 2/(1+g)$. The overall 
reflection and transmission probability in this case can be obtained by 
considering the infinite sequence of reflection and transmission from the 
two boundaries of the finite length interacting region, and are given by,
\be r(\omega) = r_0 + t_0 t_0' \sum_{n=1}^{\infty} (r_0' e^{2i \omega L/
\tilde{v}})^{2n}~ = r_0 \frac{ 1- e^{4i \omega L/\tilde{v}}}{1 - r_0^2 
e^{4i \omega L/\tilde{v}}}~, \ee
which is identical to Eq.~\eqref{chk1}. Note that $r_0 = \lambda/2(v 
+\tilde{v})$. The overall transmission coefficient 
is given by the sum of the following infinite series, 
\be t(\omega) = t_0 t_0' e^{2i \omega L/\tilde{v}}\sum_{n=0}^{\infty} 
(r_0' e^{2i\omega L/\tilde{v}})^{2n}~ = \frac{ t_0 t_0' e^{2i \omega L/
\tilde{v}}}{1 - r_0^2 e^{4i \omega L/\tilde{v}}}~, \ee
and is identical to Eq.~\eqref{chk2}. 

We thus see that the AC scattering coefficients encode the full history 
of the trajectory of the electron including multiple charge fractionalization 
events at the FL-TLL interfaces, and at the junction. 

\subsection{TRS violating (${\mathbb M}_1$) fixed points}
We now consider the case of $Y$-junctions which do not preserve TRS, 
{\it i.e.~} the ${\mathbb{M}}_1$ class of fixed points. In this 
case the ${\mathbb S}$ matrix has the same cyclic form of the 
${\mathbb{M}}_{1}$ class of matrices and it has only three independent 
elements, since all the diagonal elements of ${\mathbb{M}}_{1}$ are identical. 
Following the same procedure as in the previous case, we calculate the elements of $\mathbb{S}$ to be 
\begin{widetext}
\ber 
s_{11}& = & \lambda^{-1} \eta^{-1} \Big[\tilde{v} \Big\{8 
\tilde{v} \Big(2 \tilde{v} e^{3 i k L} \cos (k L) (2 \lambda \cos \theta +3 
\cos (2 k L) (-2 \lambda \cos \theta -\lambda +4v) + \lambda -12 v) -3 
\tilde{v}^2 \left(-1+e^{2 i k L}\right) \nn \\ 
& \times & \left(1+e^{2 i k L}\right)^2 + i e^{3 i k L} \sin (k L) (3 \lambda 
(2 \cos \theta +1) (2v - \lambda) \cos (2 k L)-2 \lambda \cos \theta (2 v+
\lambda )+(6 v+\lambda ) (4v-\lambda ))\Big) \nn \\
&-& 3 \left(-1+e^{2 i k L}\right)^2 \left(1+e^{2 i k L}\right)
\left(\lambda ^3+16 v^3-8 \lambda v^2 \cos \theta -4 \lambda v^2\right)\Big\} 
\nn \\ 
&+& 3 v \left(-1+e^{2 i k L}\right)^3 (2v-\lambda) \left(\lambda ^2+4 v^2-4
\lambda v \cos \theta \right)\Big] ~,
\eer
where
\be
\eta ~=~ 12 ~e^{3 i k L} ~[2\zeta - \lambda \sin(kL) ] \times \left( [2\zeta 
- \lambda \sin(kL) ]^2 + 4 i \lambda \sin (kL) \zeta (1-\cos \theta) \right)~, 
\ee
\end{widetext}
and
\be 
\zeta = \tilde v \cos (kL) - i v \sin(kL)~. 
\ee
The other elements of $\mathbb{S}$ are given by 
\ber
s_{21}&=& -48 ~\eta^{-1}~ \tilde{v}^2 e^{3 i k L} \left[ 2 \zeta 
c \right. -\left. i 
\lambda \sin(kL) b ~
\right]~, \\ 
s_{31}&=& 48 ~\eta^{-1}~ \tilde{v}^2 e^{3 i k L} \left[ 2 \zeta^* 
c + i \lambda \sin(kL) 
b \right]~, 
\eer
along with $s_{22}=s_{33}= s_{11},~ s_{12}=s_{23}= s_{31}$, and 
finally $s_{13}=s_{32} =s_{21}$.

In Fig.~\ref{fig1}, we plot the absolute values of some of the elements of 
$\mathbb{S}$, as 
a function of the incoming energy ($\omega=\tilde{v} k$) and the parameter 
$\theta $ describing the fixed points of the junction. Note that unlike the 
DC conductivity, the AC current amplitudes carry signatures of the 
{\it e-e} interactions, {\it i.e.~}they depend on the {\it e-e} interaction strength 
$\lambda$, and the finite length $L$ of the TLL wires. The amplitudes 
oscillate as a function of the frequency of the incident AC current with 
a period of $ 2 \pi \tilde{v}/L$ for the ${\mathbb{M}}_1$ class of fixed 
points and with a period of $ \pi \tilde{v}/L$ for the ${\mathbb{M}}_2$ class 
of fixed points as can be seen from Fig.~\ref{fig1} (a) and \ref{fig1} (c) 
respectively. 
Experimentally such measurements of oscillations of the AC current amplitude, 
as a function of the frequency may be used to classify the $Y$-junctions, 
whose fixed point may not be known {\it a priori}. 

\begin{figure}[t]
\begin{center} 
\includegraphics[width=1.0 \linewidth]{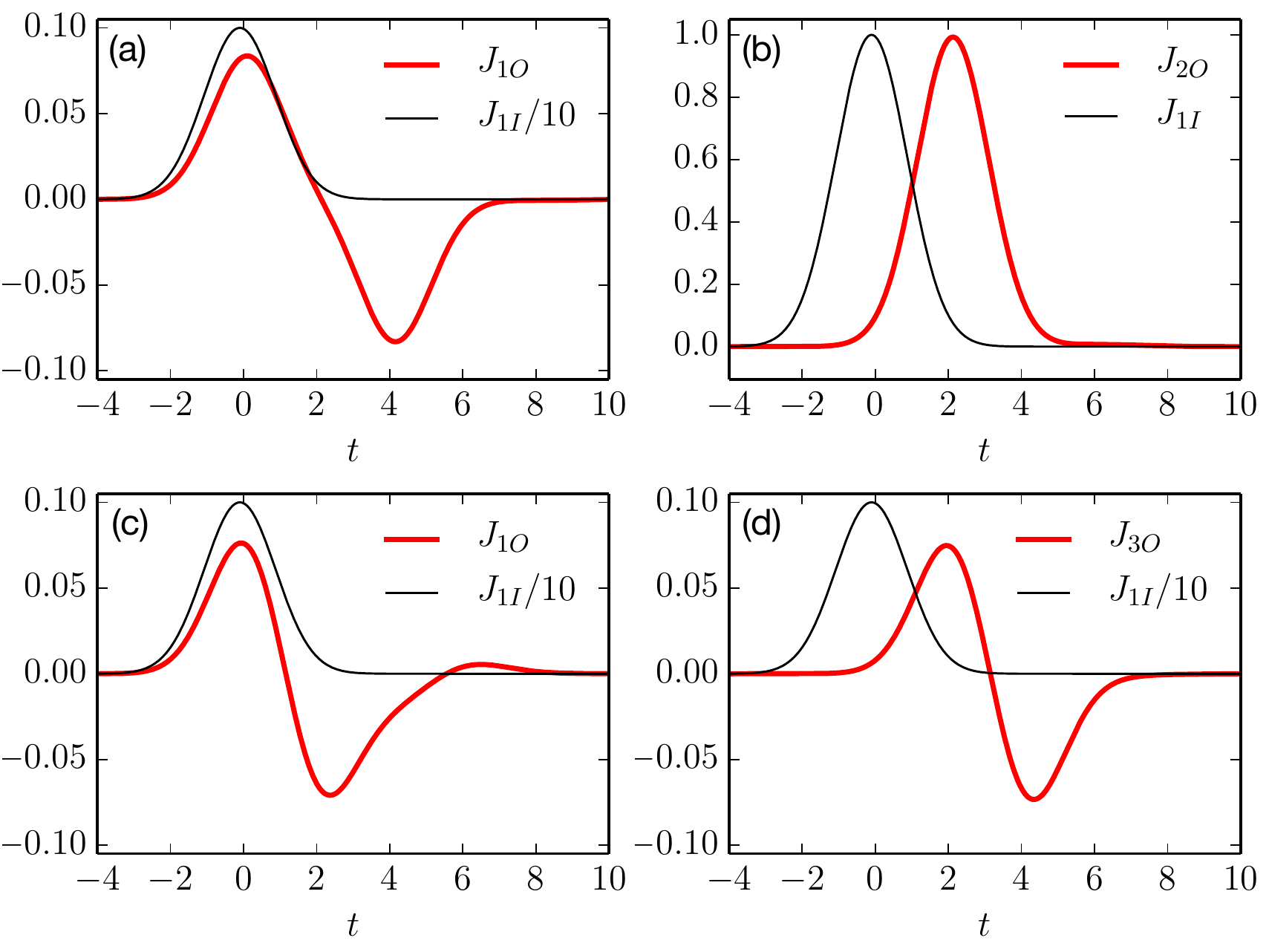}
\end{center}
\caption{Pulse propagation in a $Y$-junction as a function of time, for an 
incoming current only in wire $1$. For the junction tuned to $\theta = 0$ fixed 
point of class ${\mathbb{M}}_2$, panel (a) shows the outgoing reflected 
current in wire $1$, and panel (b) shows the outgoing transmitted current in 
wire $2$. 
In this case, wires $1$ and $2$ are completely connected effectively becoming 
one wire, and wire $3$ is completely disconnected. For the junction tuned to 
the chiral fixed point $\chi_{+}$, {\it i.e.~}$\theta = 4 \pi/3$ of the 
${\mathbb{M}}_1$ class of fixed points, panel (c) shows the outgoing 
(reflected) current in wire $1$, and panel (b) shows the outgoing 
(transmitted) current in wire $3$. 
\label{fig2} }
\end{figure}

Motivated by recent time-resolved experiments on 1D TLL wires 
[\onlinecite{Kamata, Perfetto2014}], we study the propagation of a 
wave packet incident from FL lead in wire $1$, in Fig.~\ref{fig2}. 
Note that the results depicted in Fig.~\ref{fig2} (a) are similar to the
results for reflected current in type-I geometry for a 1D TLL wire, 
reported in Ref. [\onlinecite{Kamata}]. Our results generalize the recent 
results of Ref.~[\onlinecite{Perfetto2014}], for arbitrary shaped wave packet 
propagation in a single TLL wire to the case of multi-wire junctions.

\subsection{ The DC limit}
In the linear response regime 
the DC conductivity of the $Y$-junction is different if the TLL wires
are connected to FL leads and if the TLL wires extend to infinity. This is 
well known for the case of a single TLL wire, whose linear DC 
conductance is $e^2/h$ when connected to FL leads and is 
$ge^2/h$ for an infinite TLL wire [\onlinecite{Stone}]. 

To obtain the DC conductivity for a $Y$-junction connected to FL leads, 
from our AC results, we note that in the DC limit, {\it i.e.~}as 
$\omega ~\to~ 0$ (or as $k \to 0$), for both classes of fixed points we have 
\be 
\lim_{\omega \to 0}{\mathbb S} = {\mathbb{M}}~.
\ee
This implies $ I^{\rm out}_i = \sum_j {\mathbb{M}}_{ij} I^{\rm in}_j$, where we have defined $I^{{\rm in}({\rm out})}_i$ to be the current flowing towards 
(away from) the junction in the TLL wire $i$.
Further if $V_j$ is the voltage applied in the FL lead connected to wire $j$, 
then the incoming current (per spin) is 
related to it by $I^{\rm in}_i = \sum_j (e^2/h)~{\delta}_{ij} V_j$. Now 
using the definition of the junction conductance $\mathbb{G}$, which relates the net current flowing towards the junction to the external voltages, {\it i.e.~}$I_i \equiv I^{\rm in}_i - I^{\rm out}_i= \sum_j \mathbb{G}_{ij} V_j$, 
we obtain the DC conductance matrix (per spin orientation) to be 
\be {\mathbb{G}} =(e^2/h) ({\mathbb{I}}~-~{\mathbb{M}})~. \label{GF} \ee

For a $Y$-junction with TLL leads extending to infinity, the voltage applied 
in the LL lead of wire $j$ is related to the incoming current by 
$I^{\rm in}_i = \sum_j (g e^2/h)~{\delta}_{ij} V_j$, and the current splitting 
matrix at the junction is $\tilde{\mathbb{M}}$. Thus the conductance matrix 
(per spin) is given by 
\be {\mathbb{G}} \label{GL}~=~ 
(g e^2/h) ({\mathbb{I}}~-~\widetilde{\mathbb{M}})~. 
\ee
As a check of Eqs.~\eqref{GF} and \eqref{GL}, we note that they are consistent 
with the conductance of several fixed points reported in 
Ref.~[\onlinecite{chamon1}] using the Kubo formula and other methods.
We emphasize here that the DC conductivity for a junction of finite length TLL wires
connected to FL leads does not carry any 
signature of interactions and charge fractionalization events in the system. 
In contrast, the AC conductivity depends on the {\it e-e} interactions as well 
as the length of each wire.

In the next section, we consider a $Y$-junction of infinite length TLL 
wires --- [see Fig.~\ref{fig0} (b)], with point-like tunneling impurities at 
the junction, and calculate the `tunneling' current
and quantum noise at the junction.

\section{Tunneling current and tunneling noise at the junction}
\label{secIV}

We now consider the effect of point-like 
charge conserving tunneling operators between infinite TLL wires, at the 
junction ($x=0$), and study the tunneling current and low frequency quantum 
noise arising due to 
these.  Note that each of the boundary condition at the junction characterized by $\theta$ corresponds
to a scale invariant boundary condition, or a RG fixed point,  of the bosonic field theory. 
The knowledge of $\mathbb{M}$ at each $\theta$ thus completely specifies 
all the reflection and transmission amplitudes at the level of the Hamiltonian for the Y-junction --- see Ref.~[\onlinecite{chamon1}]. 
Additional small tunneling (boundary operators) between wires may be treated as a small variation 
of the amplitudes. 
If all of the tunneling operators at the junction, are irrelevant in a RG sense, then the fixed point is stable, otherwise switching on of relevant tunneling operators around any fixed point makes the junction `flow' to another fixed point on changing length and energy scales in the system. In our case, a tunneling operator is relevant (irrelevant),  if the boundary scaling dimension of 
the tunneling operators is less than (greater than) unity, {\it i.e.~}$d_0 < 1$  ($d_0 > 1$). However we emphasize that, as long as the wire length is not very large (as compared to the other length scales set by the temperature or the external voltages), such that the RG flow does not take one far away from the fixed point (either stable or unstable),  the calculations described in this section are still valid for all fixed points.  
 A similar set-up was used in 
Ref.~[\onlinecite{Feldman_PRB2011}]
to study rectification in a $Y$-junction of TLL wires, which was found to be 
strongest for junctions violating TRS, and for strongly coupled junctions.




We consider very narrow (point-like) tunneling barriers, so that the time 
duration of the tunneling event is much smaller that the time duration between 
two successive tunneling events. Such discrete tunneling events lead to the so 
called `shot noise', whose spectrum carries the signature of correlations 
between different tunneling events. In addition, we also have different 
voltages in different leads. This leads to the so called `Josephson noise', 
which arises from the quantum interference of the wave-functions, on different 
sides of the tunneling impurity (different wires in our case), and it may 
lead to a divergence 
in the noise spectrum at frequency $\omega = q V_{\rm eff}/h$, where 
$V_{\rm eff}$ is the effective voltage difference that the tunneling operator is 
subjected to [\onlinecite{chamon_prb1995}]. In what follows,  we derive 
the tunneling noise at the junction from a perturbative calculation, which 
gives both the shot noise and Josephson noise contributions.

In a clean junction (no tunneling `impurity'), the current in wire $i$ is 
given by $I_i = {\mathbb G}_{ij} V_j$, where ${\mathbb G}$ is given by 
Eq.~\eqref{GL} for a $Y$-junction with infinite TLL wires. The switching on of 
tunneling operators ($\psi_{iO}^\dagger \psi_{jI}$) in the vicinity 
($x_i \le1/k_{\rm F}$ ) of the $Y$-junction which is 
tuned to be at a particular fixed point, leads to an additional tunneling
current $\delta I_i$, such that  $I_i^{\rm total} = 
I_i + \delta I_{i}$. If the tunneling Hamiltonian is expressed as, 
\be \label{eq:Htunn} H_{\rm tunn}~=~ \gamma ~\psi_{iO}(0,t)^\dagger ~
\psi_{jI}(0,t) +h.c~, \ee
then the tunneling current operator ($\delta \hat{I}_{\rm i}$) is defined by 
\ber \delta \hat I_{i} (t) &=& q \frac{d\hat \rho_{iO}}{dt} = -i q \hbar^{-1}
\left[\hat \rho_{iO} , \hat H_{\rm tunn}\right] \\
&=& i q \gamma \hbar^{-1} [\psi^\dagger_{iO} \psi_{jI} 
- h.c.~]~. \nn \eer
It can be calculated at any time $t$ from the following expression, 
\be \langle \delta \hat I_{i} \rangle = \langle 0| ~S(-\infty;t) ~\delta 
\hat I_{i} (t)~ S(t;-\infty) ~|0 \rangle~, \ee
where $|0 \rangle$ denotes the ground state of the unperturbed system, 
{\it i.e.}~the initial state at $t \to -\infty$. Here $S$ is the scattering 
matrix arising due to the tunneling impurities, and it is given by 
\be 
S(t;-\infty) = S^\dag (-\infty;t) ={\cal T} e^{-i \hbar^{-1}
\int_{-\infty}^{t} \hat H_{\rm tunn}(t') dt' }~, \ee
where ${\cal T}$ denotes the time ordering operator.
Using the notation: $\hat B_{ij}(x,t) \equiv \psi_{iO}^\dagger 
\psi_{jI}$ for the tunneling operator, and expressing the fermonic operators 
in terms of bosonic fields using Eq.~\eqref{BI}, we get
\ber \hat B_{ij}(x,t) &=& \frac{1}{2 \pi \alpha} F_{iO}^\dagger F_{jI}~ e^{i 
(2 \pi/L)(N_{iO}-N_{jI})vt} \nn \\
& \times & e^{-i\phi_{jI}(x,t)}~ e^{-i\phi_{iO}(x,t)}~. \eer
The tunneling current, in terms of the tunneling operator is, 
$\delta \hat I_{i} (t) = i q \gamma \hbar^{-1} [\hat B_{ij}(t) - h.c.]$, 
while the scattering 
matrix is given by $S (t,-\infty) = 1 - i\hbar^{-1} \gamma \int_{-\infty}^t 
dt' [\hat B_{ij}(t') + h.c.]$, up to first order in the tunneling amplitude 
$\gamma$. Thus the expectation value of the tunneling current operator, 
up to second order in $\gamma$, is given by 
\ber \label{eq:Ibs}
\delta I_{i} &=&\frac{q\gamma^2}{ \hbar^2 } \int_{-\infty}^t dt' \times \\
&& \langle 0| ( \hat B_{ij}^\dagger (t) B_{ij}(t') - \hat B_{ij}(t') \hat 
B_{ij}^\dagger (t) ) + h.c. ~|0 \rangle ~. \nn \eer
The symmetrized noise is given by the Fourier transform of the 
current-current correlator:
\be S(\omega)= \int_{-\infty}^{\infty} ~dt~ e^{-i\omega t} \langle \delta 
\hat I_{i}(t)\delta \hat I_{i}(0) +\delta \hat I_{i}(0)\delta \hat I_{i}(t)
\rangle~, \ee
and up to second order in $\gamma$, we obtain, 
\ber S(\omega) &=& \frac{q^2\gamma^2}{\hbar^2 } \int_{-\infty}^{\infty} ~dt~ 
e^{-i\omega t} \\
&\times & \langle0| (\hat B_{ij} (t) B_{ij}^\dagger(0) + \hat B_{ij}(t) 
\hat B_{ij}^\dagger (0))+H.C. |0\rangle. \nn \eer 

\begin{table}[t!] 
\begin{center} 
\begin{tabular}{l r } \hline \hline \\
Operators (${\mathbb{M}}_1$ class) & Scaling dimension ($d_0$) \\ \hline
$\psi_{iO}^\dagger \psi_{iI}$ & $\frac{4 g (1 - \cos \theta )}{3 \left(g^2+
 \left(g^2-1\right) \cos \theta +1\right)}$ \\
$\psi_{2O}^\dagger \psi_{1I},\psi_{3O}^\dagger \psi_{2I},\psi_{1O}^\dagger 
 \psi_{3I}$ & $ \frac{2 g \left(\cos \theta +\sqrt{3} \sin \theta
 +2\right)}{3 \left(g^2+\left(g^2-1\right) \cos \theta +1\right)}$ \\
$\psi_{1O}^\dagger \psi_{2I},\psi_{2O}^\dagger \psi_{3I},\psi_{3O}^\dagger 
 \psi_{1I}$ & $\frac{2 g \left(\cos \theta -\sqrt{3} \sin \theta
 +2\right)}{3 \left(g^2+\left(g^2-1\right) \cos \theta +1\right)} $\\ 
\\ \hline \hline \\
Operators (${\mathbb{M}}_2$ class) & Scaling dimension ($d_0$)\\ \hline
$\psi_{1O}^\dagger \psi_{1I}$ & {\footnotesize $\frac{1}{3} g (2+\cos \theta -
\sqrt{3} \sin \theta)$ }\\
$\psi_{2O}^\dagger \psi_{2I}$ & {\footnotesize$\frac{1}{3} g (2+\cos \theta +
\sqrt{3} \sin \theta)$} \\
$\psi_{3O}^\dagger \psi_{3I}$ & {\footnotesize$\frac{2}{3} g (1-\cos 
\theta )$} \\
$\psi_{1O}^\dagger \psi_{2I}$,$~\psi_{2O}^\dagger \psi_{1I}$ & 
{\footnotesize$\frac{3+g^2}{6g} (1-\cos \theta )$} \\
$\psi_{2O}^\dagger \psi_{3I}$,$~\psi_{3O}^\dagger \psi_{2I}$ & 
{\footnotesize$\frac{3+g^2}{12 g}(2+ \cos \theta -\sqrt{3} \sin \theta )$ }\\
$\psi_{3O}^\dagger \psi_{1I}$,$~\psi_{1O}^\dagger \psi_{3I}$ & 
{\footnotesize$ \frac{3+g^2}{12g} (2+\cos \theta+\sqrt{3} \sin \theta )$} \\ 
 \\ \hline \hline
\end{tabular} 
\end{center}
\caption {Scaling dimensions of various tunneling operators for both 
${\mathbb{M}}_1$ and ${\mathbb{M}}_2$ classes of fixed points.}
\label{T1}
\end{table}

\begin{figure}[t]
\begin{center} 
\includegraphics[width=1.0 \linewidth]{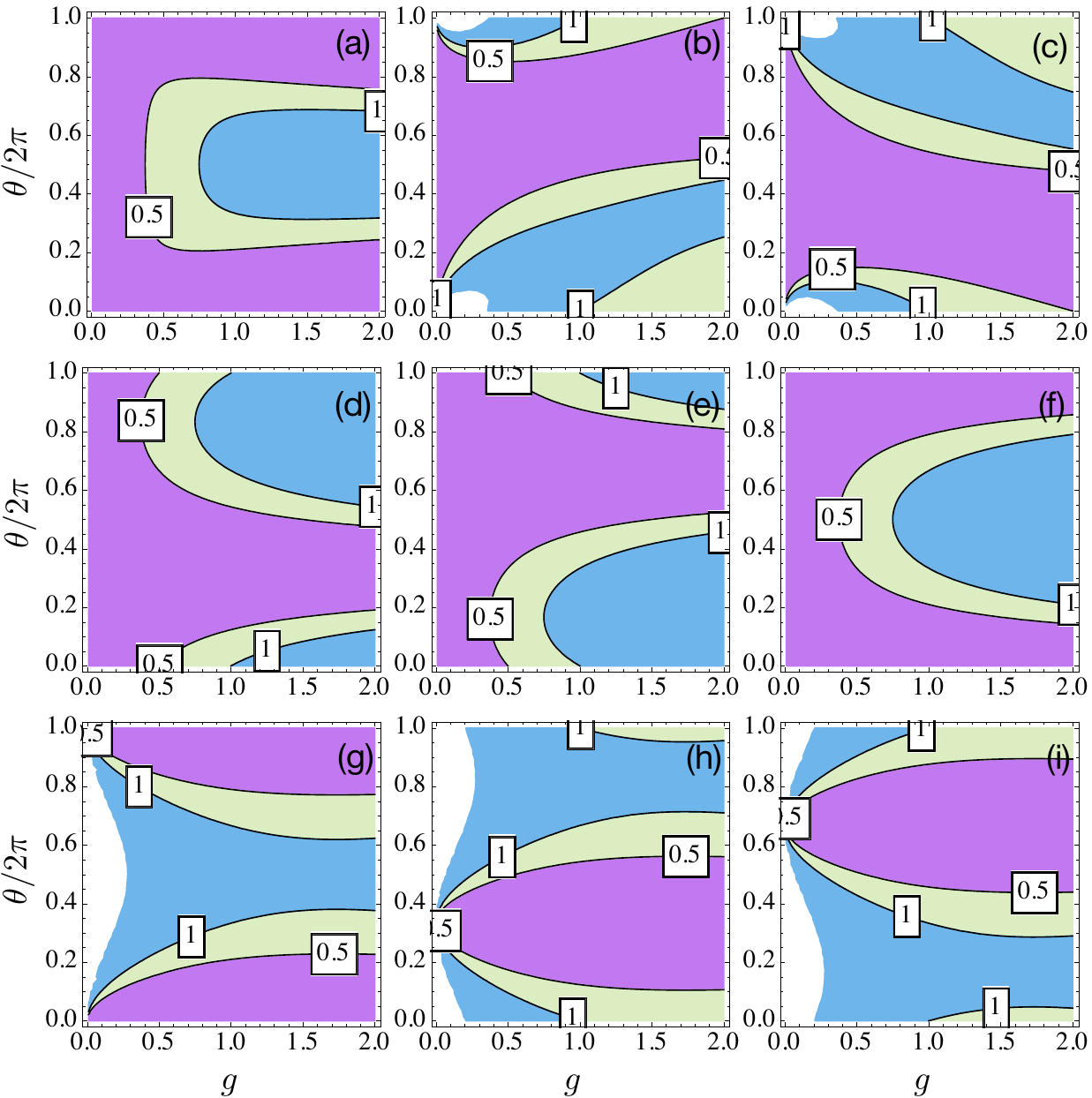}
\end{center}
\caption{Color plot of the scaling dimension for various tunneling operators. 
Panels (a), (b) and (c) represent the operators \textcolor{blue}{$\psi_{iO}^\dagger 
\psi_{iI}$, $\psi_{1O}^\dagger \psi_{2I}$ and $\psi_{2O}^\dagger \psi_{1I}$ }
respectively, for the ${\mathbb{M}}_1$ class of fixed points (as they appear 
in Table \ref{T1}). Panels (d) - (i) display the scaling dimensions of various 
tunneling operators as they appear in Table \ref{T1}, for the ${\mathbb{M}}_2$ 
class of fixed points. The purple region in all the panels has $d_0 <1/2$, the 
green region represents $1/2 < d_0< 1$ and the blue region in all the panels 
has $d_0 > 1$. Both the backscattering current and the quantum noise show 
diverging behavior for $d_0 < 1/2$, {\it i.e.}~in the $(\theta, g)$ parameter 
space represented in purple. Also note that the tunneling operators become 
relevant in the region $d_0 <1$, {\it i.e.}~purple and green regions, and 
will make the junction `flow' to another fixed point.
\label{fig3}}
\end{figure}

To obtain the final expressions for the tunneling current and for the 
symmetrized quantum noise, we need the ground state expectation values of 
operators, such as ${\cal O} \equiv \hat B_{ij}(x,t) 
\hat B_{ij}^\dagger(x',t')$. Following a standard procedure 
[\onlinecite{delft}], at zero temperature ($T$), these are given by 
\be \label{eq:expectation}
\langle 0|{ \cal O} |0\rangle = \frac{\alpha^{2d_0}}{4 \pi^2 \alpha^2} 
\frac{e^{i \frac{2 \pi}{L} \langle 0| N_{iO}-N_{jI}|0\rangle v (t-t') }}{[
(x-x')^2 - (v(t-t') - i \alpha)^2]^{d_0}} ~, \ee
where $d_0$ is the boundary scaling dimension of the tunneling operator 
involved, {\it i.e.}~$\hat B_{ij} = \psi_{iO}^\dagger \psi_{jI}$. For all 
possible tunneling operators, $d_0$ is tabulated in Table \ref{T1} for both 
the ${\mathbb{M}}_1$ and ${\mathbb{M}}_2$ class of fixed points, 
[\onlinecite{agarwal_tdos}]. 
In addition we also have terms like $N_{iO}-N_{jI}$ in the exponential whose 
expectation values depend on the external chemical potentials $\mu_i$ (or 
voltages $V_i$) applied on each wire in the grand canonical ensemble picture. 
The outgoing $N_{iO}$ are related by the current splitting matrix 
${\mathbb{M}}$ to the incoming $N_{iI}$ which are in turn 
related to the external reservoir voltages. Thus we have 
\be N_{iO}=\sum_p {\mathbb{M}}_{ip} N_{pI}~, \quad {\rm and} \quad 
\frac{h v }{L} \langle N_{iI}\rangle = \mu_{i} = q V_i ~. \ee
The expectation value of $\langle N_{iO}-N_{jI} \rangle$ now defines a 
new frequency scale which is related to external voltages by
\be \label{eq:omega} \omega_0 \equiv \frac{2\pi v}{L} \langle N_{iO}-N_{jI} 
\rangle = h^{-1}q \left(\sum_p( {\mathbb{M}}_{ip} V_p) - V_j\right)~, \ee
where $j$ and $p$ are wire indices. Physically $\hbar \omega_0 $ is the 
effective voltage difference that the tunneling operator `feels' (is subjected 
to) for an electron incoming in lead $j$ and finally outgoing in lead $i$. 

We now proceed to calculate the tunneling current by substituting 
Eqs.~\eqref{eq:omega} and \eqref{eq:expectation}, in Eq.~ \eqref{eq:Ibs}. 
A straightforward calculation, using the integral, 
\be
I_{\pm} = \int_{-\infty}^{\infty} dt' \frac{e^{\pm i \omega_0 t'}}{(
\frac{\alpha}{v} - i t')^{2 d_0} } = \frac{2 \pi~|\omega_0|^{2d_0-1}}{\Gamma
(2 d_0)} e^{-\frac{\alpha |\omega_0|}{v}} \theta(\mp \omega_0)~, \ee
gives, 
\be \label{dI}
\delta I_{i}=q \frac{2\pi\gamma^2}{h^2 \alpha^2}~\frac{1}{\Gamma(2d_0)}~
\left(\frac{\alpha}{v}\right)^{2d_0}|\omega_0|^{2d_0-1} {\rm sign}(\omega_0). 
 \ee
Here $\Gamma(2d_0)$ appearing in the denominator is the Gamma function.
The 
scaling dimension $d_0$ in general depends on the strength of interactions 
and the fixed point ($\theta$) that the junction is tuned to. It is tabulated 
in Table \ref{T1}, and a contour plot of $d_0$ in the ($\theta, g$) parameter 
space is presented in Fig.~\ref{fig3}. For the case of a `non-interacting' 
junction, {\it i.e.~}$d_0 \to 1$ (which is equivalent to the case of 
$g \to 1 $ in a single wire scenario), $ \delta I_i |_{d_0 \to 1}= q 
\frac{2\pi\gamma^2}{h^2 v^2} \omega_0$. In the limiting case of $d_0 \to 1/2$ 
(which is equivalent to the case of $g \to 1/2 $ in the single wire scenario), 
we have $\delta I_i|_{d_0 \to \frac{1}{2}} = q \frac{2\pi\gamma^2}{h^2 
\alpha v} {\rm sign}(\omega_0)$. 

To relate it to an earlier work let us consider the fixed point $\theta = 0$ 
of the ${\mathbb{M}}_2$ class, {\it i.e.}, $[a, b, c] = [1,0,0]$, which 
represents the specific case of wires $1$ and $2$ being directly connected 
and the wire $3$ being completely disconnected. Now consider a tunneling 
operator $\psi_{2O}^\dagger \psi_{2I}$, for which $\omega_0 = h^{-1}q (V_1 
- V_2)$ and $d_0 = g$. The tunneling current in this case is given by 
\be \delta I_{i}=q \frac{2\pi\gamma^2}{h^2 \alpha^2}~\frac{1}{\Gamma(2g)}~
\left(\frac{\alpha}{v}\right)^{2g}|h^{-1} q (V_1-V_2)|^{2g-1} ~, \ee
which has earlier been reported in the context of current enhancement by a 
tunneling impurity in Ref.~[\onlinecite{feldman_prb2003}], and as a 
limiting case of two or more impurity scattering in TLL wires in 
Refs.~[\onlinecite{makogon, agarwal_prb2007}]. 

Equation~\eqref{dI} can be generalized to finite temperatures by using the 
following  transformation [\onlinecite{chamon_prb1995}]:
\be \label{cm}
I_{\pm} = \int_{-\infty}^{\infty} dt' \frac{e^{\pm i \omega_0 t'}}{\left(
\frac{\alpha}{v} - i t'\right)^{2 d_0} } \to e^{i \pi d_0} 
\int_{-\infty}^\infty dt' \frac{e^{\pm i \omega_0 t'}}
{\left|\frac{\sinh(\pi T t')}{ \pi T}\right|^{2 d_0} }~, \ee
which gives, 
\be
I_{\pm} (T)= 2 (\pi T)^{2 d_0-1} B\left(d_0 + \frac{i \omega_0}{2 \pi T}, 
d_0 - \frac{i \omega_0}{2 \pi T}\right)~ e^{\pm \frac{ \omega_0}{2T}}~, \ee
where $T$ denotes the temperature in units of $k_{\rm B}/\hbar$, with 
$k_{\rm B}$ being the Boltzman constant, and $B(x,y) = B(y,x)$ is the 
$\beta$-function. The $\beta$-function can also be written in terms of the 
Gamma 
function: $B(x,y) = \Gamma(x) \Gamma(y)/\Gamma(x+y)$. The tunneling current 
at finite $T \le \hbar v / \alpha$ is now given by, 
\ber
\delta I_i (T) &=& q \frac{4\gamma^2}{h^2 \alpha^2} \left(\frac{\alpha}{v}
\right)^{2d_0} (\pi T)^{2 d_0-1} \\
&\times & B\left(d_0 + \frac{i \omega_0}{2 \pi T}, d_0 - \frac{i \omega_0}{2 
\pi T}\right) \sinh\left(\frac{ \omega_0}{2T}\right) ~ \nn. \eer
In the limiting case of $d_0 \to 1$, we can use the identity $\Gamma(1+ix) 
\Gamma(1-ix) = \pi x /\sinh(\pi x)$, to obtain $\delta I_i (T)|_{ d_0 \to 1} 
= q \frac{2\pi\gamma^2}{h^2 v^2} \omega_0$, independent of the temperature. 
For the case of $d_0 \to 1/2$, one can use the identity $\Gamma(1/2+ix) 
\Gamma(1/2-ix) = \pi /\cosh(\pi x)$, to get $\delta I_i (T)|_{d_0 \to 
\frac{1}{2}} = q \frac{4\pi\gamma^2 \alpha}{h^2 v} \tanh[\omega_0/(2T)] $. 

The symmetrized quantum noise, up to second order in the tunneling strength 
$\gamma$, can also be calculated in a similar fashion and is given by 
\ber 
S(\omega)&=& q^2 \frac{2\pi\gamma^2}{h^2 \alpha^2}~\frac{1}{\Gamma(2d_0)}~
\left(\frac{\alpha}{v}\right)^{2d_0} \nn \\ 
&\times & \left(|\omega-\omega_0|^{2d_0-1}+ |\omega +\omega_0|^{2d_0-1}\right)
\label{noise1} ~. 
\eer
It can be expressed in terms of the tunneling current as 
\be S(\omega) = q\delta I_{i}\left(|1-\omega/\omega_0|^{2d_0-1}+|1+\omega/
\omega_0|^{2d_0-1}\right) ~. \ee
%
As a check of our calculations, we note that for the specific case of 
$\theta =0$, discussed in the previous paragraph, Eq.~\eqref{noise1} of our 
manuscript, reproduces Eq.~(17) of Ref.~[\onlinecite{chamon_prb1995}] in which 
the authors studied the perturbative noise for a small point impurity in an 
otherwise clean TLL. In the limit $|\omega/\omega_0| \to 0$, or at small 
frequencies, $S(\omega) \approx 2q \delta I_{i}$ independent of 
the interaction parameter, which is the typical Schottky's shot noise result.
It corresponds to the uncorrelated arrival of particles at the tunnel barrier, whereby the time 
interval between arrival times is described by a Poissonian distribution. 
In the opposite limit of $|\omega_0/\omega| \to 0$, 
we get $S(\omega) \approx 2q \delta I_{i} |\omega/\omega_0|$, consistent with 
results for non-interacting electrons [\onlinecite{chamon_prb1995}]. In the 
limiting case of $d_0 \to 1$, for low frequencies ($\omega < \omega_0$) we 
have $S(\omega)|_{d_0 \to 1} = 2 q \delta I_i$, while for high frequencies 
($\omega > \omega_0$), we have $S(\omega) = 2 q \delta I_i \frac{\omega}{\omega_0}$ 
giving a linear dependence on the frequency. Note that the high frequency 
limit of the noise spectrum for $d_0 \to 1$, is primarily determined by 
zero point fluctuations and is independent of the applied voltages, as 
expected [{\onlinecite{Buttiker}]. In the limit $d_0 \to 1/2$, we obtain 
$S(\omega)|_{d_0 \to \frac{1}{2}} = 2 q \delta I_i$. 

\begin{figure}[t]
\begin{center} 
\includegraphics[width=.98 \linewidth]{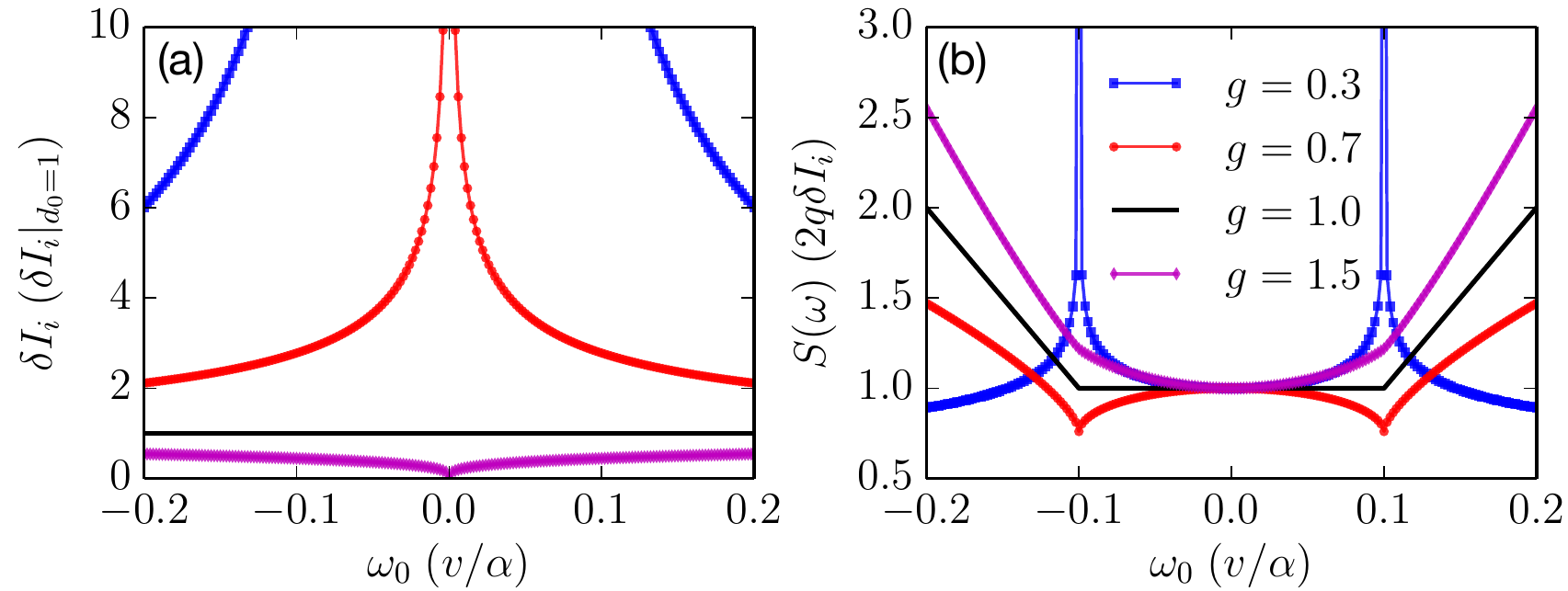}
\end{center}
\caption{ Plot of the the tunneling current in panel (a) and the quantum 
noise in panel (b) in wire $1$, as a function of $\omega_0$, for different 
values of the interaction strength. The junction is tuned to the `chiral' 
fixed point ($\chi_+$), {\it i.e.~} $\theta = 4 \pi/3$ in the $\mathbb{M}_1$ 
class, and the tunneling operator is chosen to be the backscattering operator 
$\psi_{O1}^\dagger \psi_{I1}$. The scaling dimension of $\psi_{O1}^\dagger 
\psi_{I1}$ at the $\chi_+$ fixed point for $g = (0.3, 0.7, 1, 1.5)$, which 
corresponds to the lines represented by the blue square, red circle, black 
solid, and magenta diamond markers, respectively in both the panels, is given 
by $d_0 = (0.39, 0.84, 1, 1.14)$. We have chosen $\omega = 0.1 \alpha /v$ for 
panel (b).
\label{fig4} }
\end{figure}
%

The noise power spectrum in Eq.~\eqref{noise1} can also be generalized to 
include finite temperature effects. Using the mapping of 
Eq.~\eqref{cm}, we obtain the finite temperature symmetrized noise to be 
\be \label{SwT}
S(\omega) = q^2 \frac{4\gamma^2}{h^2 \alpha^2} \left(\frac{\alpha}{v}
\right)^{2d_0} (\pi T)^{2 d_0-1} \left[ f(\omega + \omega_0) + f(\omega - 
\omega_0) \right]~, \ee
where 
\be 
f(x) = \cosh \left[\frac{x}{2 T}\right] B\left(d_0 + \frac{i x}{2 \pi T}, 
d_0 - \frac{i x}{2 \pi T}\right)~. \ee
Note that finite temperature smears the singularities of the noise power 
spectrum. The zero frequency limit of Eq.~\eqref{SwT}, gives
\be
S(\omega \to 0) = 2q\delta I_{i} (T)\coth\left[\frac{\omega_0}{2T}\right]~,
\ee
which is the equivalent of the equilibrium Johnson-Nyquist noise for a 
$Y$-junction. For the case of $d_0 \to 1$, we get $S(\omega, T)|_{d_0 \to 1} 
= q \frac{2 \pi \gamma^2}{h^2 v^2} \left[h(\omega + \omega_0) + h(\omega - 
\omega_0)\right]$ where $h(x) = x \coth[x/(2T)]$. For $d_0 \to 1/2$, we 
have, $S(\omega, T)|_{d_0 \to 1/2} = 2q \delta I_{i} (T)\coth[\omega_0/(2T)]$.

We plot the ratio $\delta I_1/\delta I_1 (d_0 =1) $ versus $\omega_0$ in 
Fig.~\ref{fig4} (a), for the backscattering operator $\psi^\dagger_{1O} 
\psi_{1I}$ when the junction is tuned to be at the $\chi_+$ fixed point. 
The divergence of this ratio whenever $2d_0 -2 <0$ is evident. The ratio 
$S(\omega)/2 q \delta I_1$, is plotted in panel (b) of Fig.~\ref{fig4}. 
This ratio diverges whenever $2 d_0 -1 <0 $.

An important difference in the three-wire case compared to the two-wire case 
is that both $\omega_0$ and $d_0$, {\it i.e.}~the frequency of divergence in 
$S(\omega)$ as well as the power law of divergence, are in general complicated 
functions of the boundary conditions at the junction (${\mathbb{M}}$), and 
the {\it e-e} interaction 
strength. Note that the noise diverges as $\omega \to \pm \omega_0$ when 
$~d_0<1/2~$. We believe that this divergence is not a limitation of our 
$\gamma^2$ perturbation theory and it will persist even if we go to higher 
orders in $\gamma$, as in the case of a `tunneling' impurity in a single TLL 
wire [\onlinecite{chamon_prb1995}]. However, this divergence is a limitation 
of our low-energy theory,  and in a realistic experimental scenario it should be regularized by 
the highest  relevant energy scale ({\it e.g.}~temperature or the maximum external voltage)
.  This is usually achieved by replacing the ultraviolet energy cut-off, $ \hbar v/\alpha$, by  $k_{\rm B} T $ or $max[ V_1, V_2, V_3]$. 

Finally we note that the results of this section are valid for an electronic 
$Y-$junction as well as for a quasi-particle junction formed from quantum 
Hall (QH) edge states. The substitution, $q \to e$ accounts for electron 
tunneling and $q \to \nu e$ takes care of quasi-particle tunneling in QH edge 
states, where $e$ denotes the electron charge and $\nu $ is the QH filling 
fraction.

\section{Summary and conclusions}
\label{summary}
In this article we investigated the AC conductivity of a $Y$-junction formed 
from finite length TLL wires connected to FL reservoirs, based on the plasmon 
scattering approach, for injected charge pulses of 
arbitrary shapes. This formalism, gives the full spatiotemporal profile of 
the charge wave packet in all the wires, and is therefore very useful for 
analyzing time resolved transport experiments in TLL wires 
[\onlinecite{Perfetto2014, Kamata}] and their junctions.
We find that unlike the DC conductivity of a `clean' junction, the AC 
conductivity depends on the strength of the {\it e-e} interactions and the length 
of the wire. Consequently it carries signatures of charge fractionalization at the 
TLL-FL interface as well as at the junction. 
The AC conductivity also displays an oscillatory behavior as a function of 
the frequency of the incoming pulse, with the 
periodicity of $\pi \tilde{v}/L$ for the time-reversal symmetric junctions, 
{\it i.e.~} junctions characterized by ${\mathbb{M}}_2$ class of fixed points, 
and with a period of $ 2 \pi \tilde{v}/L$ for junctions which break 
time-reversal symmetry, {\it i.e.~}those characterized by the 
${\mathbb{M}}_1$ class of fixed points. The limitation of our calculation is 
that it is valid only for low AC frequencies which do not breach the 
linearization regime of each TLL wire, {\it i.e.}~$\omega < v/\alpha$.

Additionally, we consider point-like tunneling impurities at the junction of 
infinite 
TLL wires, and find the corresponding tunneling current and 
quantum noise spectrum. We explicitly show that the correlations arising from strong 
{\it e-e} interactions in TLL wires, give rise to singularities in the noise 
spectrum (calculated up to second order in $\gamma$), as a function of the frequency 
or the applied voltage. The divergence in the noise spectrum for some specific 
frequencies will possibly persist to even higher orders in 
$\gamma$, and is an artefact of the effective low-energy TLL Hamiltonian 
that we are using. In any realistic experimental scenario, the high energy 
or ultraviolet cut-off $\alpha^{-1}$ will get replaced by the other energy scales 
such as the temperature or the maximum applied voltage, 
which would cut off the divergences. Another important aspect to consider 
is that these calculations are valid only in the `tunneling' limit, until
$\gamma$ does not flow (in a RG sense) beyond the TLL bandwidth 
[\onlinecite{Kane_prl1992}], {\it i.e.} $\gamma^2 
< (\alpha (|\omega\pm \omega_0|/v)^{1-2d_0} $.
Note that similar effects have been reported in a tunneling scenario in a 
two-wire junction [\onlinecite{chamon_prb1995,feldman_prb2003}], where such 
divergences occur at very strong {\it e-e} interaction 
strength of $g< 1/2$, which is a difficult regime to probe experimentally. 
However the three-wire junction offers the possibility of being tuned (by 
means of nano-gates applied in the vicinity the junction) to various fixed 
points, where these enhancement in the tunneling current and divergence 
of the quantum noise can also occur in a very wide regime of $g$, including 
attractive {\it e-e} interaction strengths --- see Fig.~\ref{fig3}. 

We firmly believe that both of these studies, {\it i.e.}~the effects of 
pulse propagation in a $Y$-junction and `backscattering' by tunneling 
impurities at the junction, will be 
very useful for interpreting time resolved experiments 
[\onlinecite{Kamata, Perfetto2014}] in multi-wire junctions of interacting 
electrons, and in the design and fabrication of quantum circuitry in the 
future.
Experimentally, TLL wire $Y$-junctions may be fabricated using carefully 
patterned 1D wires in a 2DEG, and tuned to various fixed points by means of 
nano-gates applied near the junction. Another possibility is an `island' 
set-up proposed in Ref.~[\onlinecite{das2}], formed from quantum-Hall edge 
states, which may be more feasible. In this case, the tunneling operators 
can be controlled by means of gate voltage operated constrictions in the 
central region of the `island'. 

\section*{Acknowledgment} We thank Diptiman Sen for stimulating discussions
and for carefully reading the manuscript. We gratefully acknowledge funding 
from the INSPIRE Faculty Award by DST (Govt. of India), and from the 
Faculty Initiation Grant by IIT Kanpur, India.

\end{document}